\def\sgline{\noalign{\vskip 0.15truecm\hrule\vskip 0.15truecm}}

\def\calB{\mbox{${\cal B}$}}

\newcommand{\ra}{\rightarrow}
\newcommand{\bbbar}{\mbox{$B\overline{B}$}}
\newcommand{\DE}{\mbox{$\Delta E$}}

\newcommand{\Bomegapi}{\mbox{$B^+\rightarrow\omega\pi^+$}}
\newcommand{\Bomegak}{\mbox{$B^+\rightarrow\omega K^+$}}
\newcommand{\Bomegah}{\mbox{$B^+\rightarrow\omega h^+$}}

\newcommand{\Bomegarhop}{\mbox{$B^+\rightarrow\omega \rho^+$}}
\newcommand{\Bomegarhoz}{\mbox{$B^0\rightarrow\omega \rho^0$}}
\newcommand{\Bomegakz}{\mbox{$B^0\rightarrow\omega K^{0}$}}
\newcommand{\Bomegapiz}{\mbox{$B^0\rightarrow\omega \pi^{0}$}}
\newcommand{\Bomegakstz}{\mbox{$B^0\rightarrow\omega K^{*0}$}}
\newcommand{\Bomegakstp}{\mbox{$B^+\rightarrow\omega K^{*+}$}}

\newcommand{\Bphik}{\mbox{$B^+\ra\phi K^+$}}
\newcommand{\Bphikz}{\mbox{$B^0\ra\phi K^0$}}
\newcommand{\Bphipi}{\mbox{$B^+\ra\phi\pi^+$}}
\newcommand{\Bphipiz}{\mbox{$B^0\ra\phi\pi^0$}}

\newcommand{\Brhozpi}{\mbox{$B^+\rightarrow\rho^0 \pi^+$}}
\newcommand{\Brhozk}{\mbox{$B^+\rightarrow\rho^0 K^+$}}
\newcommand{\Brhozpiz}{\mbox{$B^0\rightarrow\rho^0 \pi^0$}}
\newcommand{\Brhozkz}{\mbox{$B^0\rightarrow\rho^0 K^0$}}
\newcommand{\Brhompi}{\mbox{$B^0\rightarrow\rho^- \pi^+$}}
\newcommand{\Brhomk}{\mbox{$B^0\rightarrow\rho^- K^+$}}

\newcommand{\Bkstzpiz}{\mbox{$B^0\rightarrow K^{*0} \pi^0$}}
\newcommand{\Bkstzpi}{\mbox{$B^+\rightarrow K^{*0} \pi^+$}}
\newcommand{\Bkstppi}{\mbox{$B^0\rightarrow K^{*+} \pi^-$}}
\newcommand{\Bkstpk}{\mbox{$B^0\rightarrow K^{*+} K^-$}}
\newcommand{\Bkstzk}{\mbox{$B^+\rightarrow K^{*0} K^+$}}

\newcommand{\omegak}{\mbox{$\omega K^+$}}
\newcommand{\omegakz}{\mbox{$\omega K^0$}}
\newcommand{\omegapi}{\mbox{$\omega\pi^+$}}
\newcommand{\omegapiz}{\mbox{$\omega\pi^0$}}
\newcommand{\omegah}{\mbox{$\omega h^+$}}

\newcommand{\omegakstz}{\mbox{$\omega K^{*0}$}}

\newcommand{\omegakstp}{\mbox{$\omega K^{*+}$}}

\newcommand{\omegarhoz}{\mbox{$\omega \rho^0$}}
\newcommand{\omegarhop}{\mbox{$\omega \rho^+$}}

\newcommand{\phik}{\mbox{$\phi K^+$}}
\newcommand{\phikz}{\mbox{$\phi K^0$}}
\newcommand{\phipi}{\mbox{$\phi\pi^+$}}
\newcommand{\phipiz}{\mbox{$\phi\pi^0$}}

\newcommand{\rhozpi}{\mbox{$\rho^0 \pi^+$}}
\newcommand{\rhozk}{\mbox{$\rho^0 K^+$}}
\newcommand{\rhozpiz}{\mbox{$\rho^0 \pi^0$}}
\newcommand{\rhozkz}{\mbox{$\rho^0 K^0$}}
\newcommand{\rhompi}{\mbox{$\rho^- \pi^+$}}
\newcommand{\rhomk}{\mbox{$\rho^- K^+$}}

\newcommand{\kstzpiz}{\mbox{$K^{*0} \pi^0$}}
\newcommand{\kstzpi}{\mbox{$K^{*0} \pi^+$}}
\newcommand{\kstzkkp}{\mbox{$K^{*0}_{K^+\pi^-} K^+$}}
\newcommand{\kstppikz}{\mbox{$K^{*+}_{K^0\pi^+} \pi^-$}}
\newcommand{\kstppikp}{\mbox{$K^{*+}_{K^+\pi^0} \pi^-$}}
\newcommand{\kstppi}{\mbox{$K^{*+} \pi^-$}}
\newcommand{\kstpkkz}{\mbox{$K^{*+}_{K^0\pi^+} K^-$}}
\newcommand{\kstpkkp}{\mbox{$K^{*+}_{K^+\pi^0} K^-$}}
\newcommand{\kstpk}{\mbox{$K^{*+} K^-$}}

\newcommand{\psfile}[3][]{ 
  \begin{center}
    \setlength{\epsfxsize}{#3\linewidth}\leavevmode
    \def\noOpt{}\def\testit{#1}\ifx\testit\noOpt%
      \epsfbox{#2}%
    \else%
      \epsfbox[#1]{#2}%
    \fi
  \end{center} 
}

\documentstyle[aps,prl,preprint,floats,epsfig]{revtex}

\begin{document}

\preprint{\tighten\vbox{\hbox{\hfil CLEO CONF 99-13}
}}

\title{Charmless Hadronic $B$ Decays to Exclusive \\
 Final States with a $K^*$, $\rho$, $\omega$, or $\phi$ Meson.}

\author{CLEO Collaboration}
\date{\today}

\maketitle
\tighten

%
%

\begin{abstract}

We present results of searches for $B$-meson decays to
charmless final states that include a $K^*$, $\rho$, $\omega$, or $\phi$
meson accompanied by a second meson. Using the entire data sample of
$9.7 \times 10^6$ \bbbar\ pairs collected with the CLEO II and CLEO
II.V detectors, we observe a signal for the decay \Bomegapi, and
measure a branching fraction of ${\cal B}(\Bomegapi) = (11.3^{+3.3}_{-2.9}
\pm 1.5) \times 10^{-6}$. We also see evidence for the decay
\Bomegakz, and set limits for the decays \Brhozpiz\ and \Bkstzpiz. In addition
to these new results, we also summarize previous CLEO results on related
channels. All quoted results are preliminary.

\end{abstract}

\newpage

{
\renewcommand{\thefootnote}{\fnsymbol{footnote}}

\begin{center}
M.~Bishai,$^{1}$ S.~Chen,$^{1}$ J.~Fast,$^{1}$
J.~W.~Hinson,$^{1}$ J.~Lee,$^{1}$ N.~Menon,$^{1}$
D.~H.~Miller,$^{1}$ E.~I.~Shibata,$^{1}$ I.~P.~J.~Shipsey,$^{1}$
Y.~Kwon,$^{2,}$
% ...
\end{center}
 
\small
\begin{center}
$^{1}${Purdue University, West Lafayette, Indiana 47907}\\
$^{2}${University of Rochester, Rochester, New York 14627}\\
% ...
\end{center}

\setcounter{footnote}{0}
}
\newpage

%
%


\section{INTRODUCTION}

 The study of charmless hadronic decays of $B$ mesons plays a key role
 in understanding the phenomenon of $CP$ violation within the Standard
 Model~\cite{ckm}. Large asymmetries are predicted for some exclusive
 final states, and although the relatively small branching fractions
 of ${\cal O}(10^{-5})$ currently limit the experimental reach for
 such studies, the sensitivity of the CLEO II detector allows us to
 measure branching fractions for some decay modes. $B$ factory
 experiments that just started or are about to begin operation should
 expand this already rich field of investigation. 

 Theoretical predictions typically make use of effective Hamiltonians, often
 with factorization
 assumptions~\cite{desh,chau,xing,dean,fl,dav,kpsvv,kps,du,ebert}. The strong
 interaction between particles in the final state complicates these
 predictions. However, experimental measurements can be used to verify the
 validity of the assumptions made, to tune the parameters of the theory in
 order to make further predictions, and to understand the relative importance
 of the various decay amplitudes that contribute to a particular
 decay. Recently, it has been suggested~\cite{rosner,desh99,hou} that
 published experimental results on charmless hadronic $B$ decays indicate that
 $\cos{\gamma} < 0$, where $\gamma$ is one of the angles of the unitarity
 triangle. This somewhat disagrees with current fits to the information most
 sensitive to CKM matrix elements~\cite{ckmfits}. Again, more experimental
 results can help clarify the situation.


 In this paper, we present preliminary results of searches for
 $B$-meson decays to exclusive two-body final states~\cite{conj} that
 include a $K^*$, $\rho$, $\omega$, or $\phi$ meson, and another
 low-mass meson. We concentrate on new results, although previously
 reported results are included for completeness.


\section{DATA SAMPLE AND INITIAL EVENT SELECTION}

 The data were collected with the CLEO II~\cite{CLEOdet} and CLEO
 II.V~\cite{CLEO2.5} detectors at the Cornell Electron Storage Ring
 (CESR). For most of the new results the data sample includes all data
 collected prior to the de-commissioning of CLEO II.V, in preparation for
 a significant upgrade. The total integrated luminosity is
 9.13~fb$^{-1}$ for the reaction $e^+ e^- \rightarrow \Upsilon(4S)
 \rightarrow B\overline{B}$, which corresponds to $9.7 \times 10^6$
 $B\overline{B}$ pairs. This is between 40\% and a factor of three
 more statistics than for previously published
 results~\cite{omega,aps99}. In addition, we re-analyzed the CLEO II
 data set with improved calibration constants and track-fitting
 procedure, allowing us to extend our geometric acceptance and track
 quality requirements. This has led to an overall increase in
 reconstruction efficiency of $10$--$20$ $\%$ compared to the
 previously published analyses. For studies of background from
 continuum processes, we also collected 4.35~fb$^{-1}$ of data at a
 center-of-mass energy below the threshold for $B\overline{B}$
 production.

 The final states of the decays under study are reconstructed by
 combining detected photons and charged pions and kaons. The
resonances in the final state are identified via the decay modes $\rho
\ra \pi\pi$, $K^* \ra K\pi$, $\omega \ra
 \pi^+ \pi^- \pi^0$, and $\phi \ra K^+ K^-$. The detector elements
 most important for the analyses presented here are the tracking
 system, which consists of several individual concentric detectors
 operating inside a 1.5 T superconducting solenoid, and the
 high-resolution electromagnetic calorimeter, made of 7800 CsI(Tl)
 crystals. For CLEO II, the tracking system consists of a 6-layer
 straw tube chamber, a 10-layer precision drift chamber, and a
 51-layer main drift chamber. The main drift chamber also provides a
 measurement of the specific ionization loss, used for
 particle identification.  For CLEO II.V the 6-layer straw tube
 chamber was replaced by a 3-layer double-sided-silicon vertex
 detector, and the gas in the main drift chamber was changed from an
 argon-ethane to a helium-propane mixture.

 Reconstructed charged tracks are required to pass quality cuts based
 on their track fit residuals and impact parameter in both the
 $r$--$\phi$ and $r$--$z$ planes, and on the number of main drift chamber
 measurements. Each event must have a total of at least four good
 charged tracks. The specific ionization ($dE/dx$) measured in the
 drift layers is used to distinguish kaons from pions. Expressed as
 the number of standard deviations from the expected value, $S_i$ ($i =
 \pi$ or $K$), it is required to satisfy $|S_i| < 3.0$. Electrons are
 rejected based on $dE/dx$ and the ratio of the measured track
 momentum and the associated shower energy in the calorimeter. Muons
 are rejected by requiring that charged tracks penetrate fewer than
 seven interaction lengths of material. Pairs of charged tracks used
 to reconstruct $K^0$s (via $K^0_S \rightarrow \pi^+ \pi^-$) are
 required to have a common vertex displaced from the primary
 interaction point. The invariant mass of the two charged pions is
 required to be within two standard deviations (10 MeV) of the nominal
 $K^0_S$ mass. Furthermore, the $K^0_S$ momentum vector, obtained with
 a mass-constrained kinematic fit of the charged pions' momenta, is
 required to point back to the beam spot.

 Photons are defined as isolated showers, not matched to any charged tracks,
 with a lateral shape consistent with that of photons, and with a measured
 energy of at least 30~(50)~MeV in the calorimeter region $|\cos{\theta}| <
 0.71 (\geq 0.71)$, where $\theta$ is the polar angle.  Pairs of photons are
 used to reconstruct $\pi^0$s. The momentum of the pair is obtained with a
 kinematic fit of the photons' momenta with the $\pi^0$ mass constrained to
 its nominal value. The resolution of the invariant mass of the two photons
 depends on the momentum of the $\pi^0$ and is between 5 and 10~MeV/$c^2$. We
 require the reconstructed mass to be within $3\sigma$ on the low side of the
 nominal $\pi^0$ mass, and $2\sigma$ on the high side.


\section{ANALYSIS TECHNIQUE}

 The primary means of identification of $B$ meson candidates is
 through their measured mass and energy. The quantity $\Delta E$ is
 defined as $\Delta E \equiv E_1 + E_2 - E_b$, where $E_1$ and $E_2$
 are the energies of the two mesons in the final state, and $E_b$ is
 the beam energy. The beam-constrained mass of the candidate is
 defined as $M \equiv \sqrt{E_b^2 - |{\bf p}|^2}$, where $\bf p$ is
 the measured momentum of the candidate. We use the beam energy
 instead of the measured energy of the $B$ candidate to improve the
 mass resolution by about one order of magnitude.

 The large background from continuum quark--anti-quark ($q\bar q$) production
 can be reduced with event shape cuts. Because $B$ mesons are produced almost
 at rest, the decay products of the $B\bar B$ pair tend to be isotropically
 distributed, while particles from $q\bar q$ production have a more jet-like
 distribution. The cosine of the angle $\theta_{S}$ between the sphericity
 axis~\cite{sphericity} of the charged particles and photons forming the
 candidate $B$ and the sphericity axis of the remainder of the event should
 have a flat distribution for $B$ mesons and be strongly peaked at $\pm1.0$
 for continuum background. We require $|\cos{\theta_{S}}| < 0.8$. For final
 states containing an $\omega$ or $\phi$ meson we use the thrust~\cite{thrust}
 axis instead of the sphericity, and we require $|\cos{\theta_{T}}| < 0.9$. We
 also form a Fisher discriminant (${\cal F}$)~\cite{bigrare} with the momentum
 scalar sum of charged particles and photons in nine cones of increasing polar
 angle around the sphericity axis of the candidate, the angle of the
 sphericity axis of the candidate with respect to the beam axis, and $R_2 =
 H_2/H_0$, the ratio of the second and zeroth Fox-Wolfram
 moments~\cite{fox}. For analyses with an $\omega$ or $\phi$ in the final
 state, the thrust axis replaces the sphericity axis, and the angle between
 the candidate's momentum vector and the beam axis replaces $R_2$.

 The specific final states investigated are identified via the
 reconstructed invariant masses of the $B$ daughter resonances.  For
 vector-pseudoscalar final states, further separation of signal events
 from combinatoric background is obtained through the use of the
 defined angular helicity state of the vector meson in the final
 state. The observable $\cal H$ is the cosine of the angle between the
 direction of the $B$ meson and the vector meson daughter decay
 direction (normal to the decay plane for the $\omega$), both in the
 vector meson's rest frame.
 
 Signal event yields for each mode are obtained with unbinned, multi-variable
 maximum likelihood (ML) fits. We also perform event counting analyses that
 apply tight constraints on all variables described above. Results for the
 latter are consistent with the ones presented below.
 For each input event, the likelihood (${\cal L}_i$) is defined as
\begin{displaymath}
{\cal L}_i = \sum_{i=1}^{m}n_i {\cal P}_i
\end{displaymath}
where ${\cal P}_i$ are the probabilities for each of the $m$
hypotheses of the fit, and $n_i$, the free parameters of the fit, are
the number of events in the overall sample for each hypothesis. The
${\cal P}_i$ are the product of the probability distribution functions
(PDFs) for each of the fit's input variables. For $N$ input events,
the overall likelihood is then
\begin{displaymath}
 {\cal L} = \frac{e^{-(\sum n_i)}}{N!} \prod_{i=1}^N {\cal L}_i,
\end{displaymath}
 where the first term takes into account the Poisson
 fluctuations in the number of events. In all cases, the fit
 includes hypotheses for signal decay modes and the dominant continuum
 background. For a few channels indicated below, we also include a
hypothesis for background from other $B$ decay modes. For all others,
we verified that this component is negligible.

 The variables used in the fit are $\Delta E$, $M$, ${\cal F}$, resonance
 masses, and ${\cal H}$ as appropriate. For pairs of final states
 differentiated only by the identity (charged pion or kaon) of one of the two
 mesons, we also use the $dE/dx$ measurement, $S_i$, for the high-momentum
 track and fit for both modes simultaneously. Correlations between input
 variables were investigated and found to be negligible.  For each decay mode
 investigated, the signal PDFs for the input variables are determined with
 fits to high-statistics Monte Carlo event samples generated with a
 GEANT~\cite{geant} based simulation of the CLEO detector response. The
 parameters of the background PDFs are determined with similar fits to the sum
 of off-resonance data and a sideband region of on-resonance data in $\Delta
 E$ and $M$. For $M$, the sideband is defined by $5.2 < M < 5.3$
 GeV/$c^2$. For $\Delta E$ the sideband varies depending on the decay mode;
 details are given in the sections below.  Sideband regions for each of the
 other input variables are also included in the likelihood fit sample.


\section{RESULTS}

 Table~\ref{results1} gives all the measurement results. Details specific to
 various final states are given in separate sections below. Results for decay
 modes with a $\phi$ meson in the final state and of the types $\rho^0 h^+$ and
 $K^{*0}h^+$, where $h^+$ is a charged pion or kaon, are based on a data
 sample of $5.8 \times 10^6$ \bbbar\ pairs. Results for final states $\rho^-
 h^+$, $K^{*+}h^-$, and \rhozkz\ are based on a sample of $7.0 \times 10^6$
 \bbbar\ pairs. Shown in the table are the signal event yield, the efficiency,
 the product of the efficiency and the relevant branching fractions of
 particles in the final state, the statistical significance of the observed
 yield in standard deviations, the branching fraction central value with
 statistical and systematic error, and the corresponding 90\% confidence level
 upper limit. The one standard deviation statistical error on the central
 value is determined by finding the values where the quantity $\chi^2 =
 -2\ln({\cal L}/{\cal L}_{\rm max})$ changes by one unit, where ${\cal L}_{\rm
 max}$ is the point of maximum likelihood.

 Systematic errors are separated into two major components. The first
 is systematic errors in the PDFs, which are determined with
 variations of the PDF parameters within their uncertainty, taking
 into account correlations between parameters. The second component is
 systematic errors associated with event selection and efficiency
 factors. These are determined with studies of independent data samples.
 For branching fraction central
 values, the systematic error is the quadrature sum of the two
 components. For upper limits, the likelihood function is integrated
 to find the yield value that corresponds to 90\% of the total
 area. This value is then increased by its systematic error, and
 the efficiency is reduced by one standard deviation of its systematic
 error when calculating the final upper limit.

\begin{table}[htbp]
\caption{Measurement results. Columns list the final states with resonance
decay modes as subscripts, event yield from the fit,
reconstruction efficiency $\epsilon$, total efficiency including secondary
branching fractions ${\cal B}_s$, statistical significance ($\sigma$),
branching fraction central value ${\cal B}$, and the corresponding
90\% confidence level upper limit. For central values the first error is
statistical and the second systematic.}
\begin{center}
\begin{tabular}{lrrrccc}
Final state & Yield(events) & $\epsilon$(\%) & $\epsilon\calB_s$(\%) &
Signif. & \calB($10^{-6})$ & 90\% UL($10^{-6}$) \\ \hline
\omegapi      & $28.5^{+8.2}_{-7.3}$ & 29 & 26 & 6.2 &
     $11.3^{+3.3}_{-2.9} \pm 1.5$ & 17   \\
\omegapiz     & $1.5^{+3.5}_{-1.5}$  & 22 & 19 & 0.6 &
     $0.8^{+1.9}_{-0.8} \pm 0.5$ & 5.8  \\
\omegak       & $7.9^{+6.0}_{-4.7}$  & 29 & 26 & 2.1 &
     $3.2^{+2.4}_{-1.9} \pm 0.8$ & 8.0  \\
\omegakz      & $7.0^{+3.8}_{-2.9}$  & 24 &  7.4 & 3.9 &
     $10.0^{+5.4}_{-4.2} \pm 1.5$ & 21   \\
\omegah       & $35.6^{+8.9}_{-8.0}$ & 29 & 26 & 7.3 &
     $14.3^{+3.6}_{-3.2} \pm 2.1$ &  21  \\
\omegarhop    & $10.8^{+6.6}_{-5.3}$ & 7.1 &  6.3 & 2.8 &
     $18^{+11}_{-9} \pm 6$ & 47   \\
\omegarhoz    & $3.7^{+6.0}_{-3.7}$  & 18 & 16 & 0.9 &
     $0.0^{+5.7 \; +2.9}_{-0.0 \; -0.0}$ & 11  \\
\omegakstp   & $1.0^{+3.6}_{-1.0}$  & 6.8 &  2.0 & 0.3 &
     $5^{+19}_{-5} \pm 6$ & 52   \\
\omegakstz   & $7.0^{+5.2}_{-3.9}$  & 14 & 8.3 & 2.3 &
     $9.1^{+6.7}_{-5.1} \pm 1.9$ & 19   \\
 & & & & & & \\
\rhozpi       & $26.1^{+9.1}_{-8.0}$ & 30 & 30   &  5.2 &
     $15^{+5}_{-5} \pm 4$ &   \\
\rhompi       & $28.5^{+8.9}_{-7.9}$ & 12 & 12   &  5.6 &
     $35^{+11}_{-10} \pm 5$ &   \\
\rhozpiz      & $3.4^{+5.2}_{-3.4}$  & 34 & 34 &     &
                            & 5.1  \\
\rhozk        & $14.8^{+8.8}_{-7.7}$ & 28 & 28   &      &
                          & 22 \\
\rhomk        & $8.3^{+6.3}_{-5.0}$  & 11 & 11   &      &
                          & 25 \\
\rhozkz       & $8.2^{+4.9}_{-3.9}$  &    & 10   &  2.7 &
                          & 27 \\
 & & & & & & \\
\kstzpi       & $12.3^{+5.7}_{-4.7}$ &    & 18   &      &
                          & 27 \\
\kstzpiz      & $0.1^{+2.8}_{-0.1}$  & 37 & 24   &     &
                                  & 4.2 \\
\kstzkkp      & $0.0^{+2.1}_{-0.0}$ &    &  18  &      &
                                  &  12  \\
 & & & & & & \\
\kstppikz     & $10.8^{+4.3}_{-3.5}$ &    &  7   &  5.2 &
     $23^{+9}_{-7}\pm{3}$ &   \\
\kstppikp     & $5.7^{+4.3}_{-3.2}$ &    & 4.1   &  2.5 &
     $20^{+15 \; +3}_{-11 \; -4}$ &   \\
~~\kstppi       &                     &    &       &  5.9 &
     $22^{+8 \; +4}_{-6 \; -5}$ &   \\
 & & & & & & \\
\kstpkkz      & $0.0^{+0.9}_{-0.0}$ &    &  7   &  0.0 &
                                 &  8  \\
\kstpkkp      & $0.0^{+1.3}_{-0.0}$ &    &  4.1   &  0.0 &
                                  &  17  \\
~~\kstpk        &                  &    &      &  0.0 &
                                  &  6  \\
 & & & & & & \\
\phipi        &                      & 54 &  27  &  0.0 &
                                 &  4.0 \\
\phipiz       &                      & 35 &  17  &  0.0 &
                                 &  5.4 \\
\phik         & $2.4^{+3.0}_{-1.9}$  & 53 & 26   &  1.3 &
     $1.6^{+1.9}_{-1.2} \pm 0.2$ &  5.9 \\
\phikz        & $4.3^{+3.2}_{-2.3}$  & 41 & 7.0  &  2.6 &
     $10.7^{+7.8}_{-5.7} \pm 1.1$ & 28  \\
\end{tabular}
\end{center}
\label{results1}
\end{table}

 Table~\ref{results2} shows the final results for each decay mode
 investigated. The third column of the table indicates whether the result is
 from this work or from an earlier analysis. For observations the final
 result is reported as a branching fraction central value, while for modes
 where the yield is not sufficiently significant we quote the 90\% confidence
 level upper limit. Also in the table are previously published theoretical
 estimates.

\begin{table}[htbp]
\caption{Final results and expectations from theoretical models.}
\vspace{0.4cm}
\begin{center}
\begin{tabular}{lcccl}
Decay mode & \calB($10^{-6})$ & Source & Theory \calB\ ($10^{-6}$)&References\\
\sgline
\Bomegapi   & $11.3^{+3.3}_{-2.9}\pm{1.5}$ & This work & $0.6-11$        &
    \cite{chau,dean,kps,du,oh,ciuchini,ali,cheng}     \\
\Bomegapiz  &          $< 5.8 $            & This work & $0.01-12$       &
    \cite{chau,dean,du,oh,ciuchini,ali,cheng}         \\
\Bomegak    &          $< 8.0$             & This work & $0.2-13$        &
    \cite{chau,dean,kps,du,oh,ciuchini,ali,cheng}     \\
\Bomegakz   &          $< 21$              & This work & $0.02-10$       &
    \cite{chau,dean,du,oh,ciuchini,ali,cheng}         \\
\Bomegah    & $14.3^{+3.6}_{-3.2}\pm 2.1$  & This work &                 & 
                        \\
\Bomegarhop &          $< 47$              & This work & $7-28$          &
    \cite{chau,dean,kpsvv,ciuchini,ali,cheng}         \\
\Bomegarhoz &          $< 11$              & This work & $0.005-0.4$     &
    \cite{chau,ciuchini,ali,cheng}                    \\
\Bomegakstp &          $< 52$              & This work & $0.9-15$        &
    \cite{chau,dean,kpsvv,ciuchini,ali,cheng}         \\
\Bomegakstz &          $< 19$              & This work & $0.3-12$        &
    \cite{chau,dean,ciuchini,ali,cheng}               \\
\Brhozpi    &    $15^{+5}_{-5} \pm 4$      & \cite{aps99} & $0.4-8$      &
  \cite{chau,ebert,dean,kps,du,oh,ciuchini,ali,cheng} \\
\Brhompi    &    $35^{+11}_{-10} \pm 5$    & \cite{aps99} & $26-52$      &
  \cite{chau,ebert,dean,du,oh,ciuchini,ali,cheng}     \\
\Brhozpiz   &          $<5.1$              & This work & $0.9-2.3$       &
  \cite{chau,ebert,du,oh,ciuchini,ali,cheng}          \\
\Brhozk     &          $<22 $              & \cite{aps99} & $0.1-1.7$    &
  \cite{desh,chau,dean,kps,du,oh,ciuchini,ali,cheng}  \\
\Brhomk     &          $<25 $              & \cite{aps99} & $0.2-2.5$    &
  \cite{desh,chau,dean,du,oh,ciuchini,ali,cheng}      \\
\Brhozkz    &          $<27 $              & \cite{aps99} & $0.04-1.7$   &
  \cite{desh,chau,dean,du,oh,ciuchini,ali,cheng}      \\
\Bkstzpi    &          $<27 $              & \cite{aps99} & $4-12$       &
  \cite{desh,chau,fl,du,oh,ciuchini,ali,cheng}        \\
\Bkstppi    & $22^{+8 \; +4}_{-6 \; -5}$   & \cite{aps99} & $1.2-19$     &
  \cite{desh,chau,dean,du,oh,ciuchini,ali,cheng}      \\
\Bkstzpiz   &          $<4.2$              & This work & $1.1-5$         &
  \cite{desh,chau,dean,du,oh,ciuchini,ali,cheng}      \\
\Bkstpk     &          $<6  $              & \cite{aps99} &              &
                                                      \\
\Bkstzk     &          $<12 $              & \cite{aps99} & $0.2-1$      &
  \cite{chau,kps,du,oh,ali,cheng} \\
\Bphipi     &          $<4.0$              & \cite{aps99} & $0.001-0.4$  &
    \cite{xing,dean,fl,kps,du,oh,ciuchini,ali,cheng}  \\
\Bphipiz    &          $<5.4$              & \cite{aps99} & $0.0004-0.2$ &
    \cite{xing,dean,fl,du,oh,ciuchini,ali,cheng}      \\
\Bphik      &          $< 5.9$             & \cite{aps99} & $0.3-18$     &
    \cite{desh,chau,dean,fl,dav,kps,du,oh,ciuchini,ali,cheng} \\ 
\Bphikz     &          $< 28$              & \cite{aps99} & $0.3-18$     &
    \cite{desh,chau,dean,fl,dav,du,oh,ciuchini,ali,cheng}     \\ 
\end{tabular}
\end{center}
\label{results2}
\end{table}


\subsection{Final states including an $\omega$ meson.}

The results for decay modes including an $\omega$ meson in the final
state were obtained with the full data sample ($9.7 \times 10^6$
\bbbar\ pairs). The final selection prior to the likelihood fit
requires $|\DE| < 200$ MeV. In Table~\ref{results1}, the final state
\omegah\ represents the sum of the \omegak\ and \omegapi\ states ($h^+
\equiv K^+$ or $\pi^+$). For $\omega K^*$ and $\omega\rho$ final states,
the resonance mass sidebands are $830 < M_{K^*} < 950$ MeV/$c^2$ and $600 <
M_{\rho} < 950$ MeV/$c^2$, respectively, and cross-feeds between those decay
modes are ignored when performing the fit. For $K^{*0} \ra K^+\pi^-$, there
are two possible assignments for the $K^+$. The choice is made based on the
$dE/dx$ information of the tracks. For $\rho^0 \ra \pi^+\pi^-$ the two charged
tracks are simply assumed to be pions.

 The maximum likelihood fit for \omegakstz\ and \omegarhoz\ includes a
 hypothesis for background from generic $B$ decays. PDFs for this hypothesis
 are obtained from a sample of Monte Carlo generated events that corresponds
 to approximately 1.7 times the size of the on-resonance data sample. The
 yield for \omegarhoz\ is entirely consistent with cross-feed from
 \omegakstz. The quoted branching fraction central value has been adjusted to
 take this into account. For the final states $\omega K^{*+}$ and $\omega
 \rho^+$, the $\pi^0$ from $K^{*+}$ or $\rho^+$ decay defines the daughter
 direction. Since the distribution of $\cal H$ is not known for these
 vector-vector final states we assume the worst case (${\cal H}^2$) when
 computing the efficiency, and we require ${\cal H}<0$ to reduce the large
 combinatoric background from soft $\pi^0$'s. This also results in much
reduced cross-feed between the two decay modes.

\begin{figure}[htbp]
\psfile{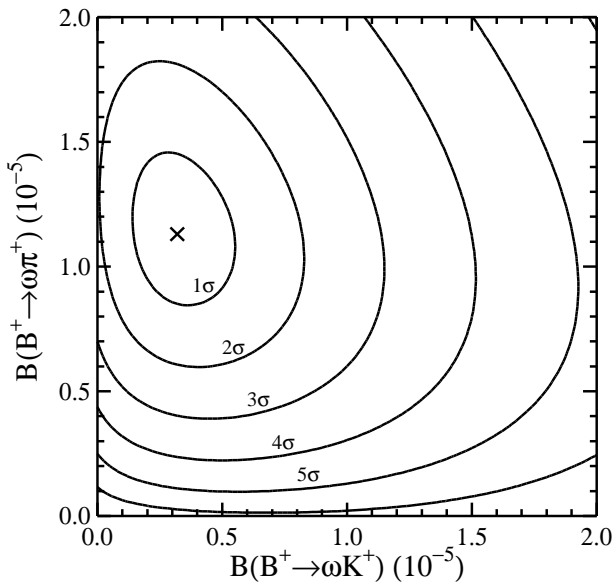}{0.5}
 \caption{\label{f:cont}
 Likelihood function contours for \Bomegapi\ and \Bomegak. Systematic
errors are not included in the contours.}
\end{figure}

\begin{figure}[htbp]
\centering
\leavevmode
\epsfxsize=3.0in
\epsfysize=3.0in
\epsffile{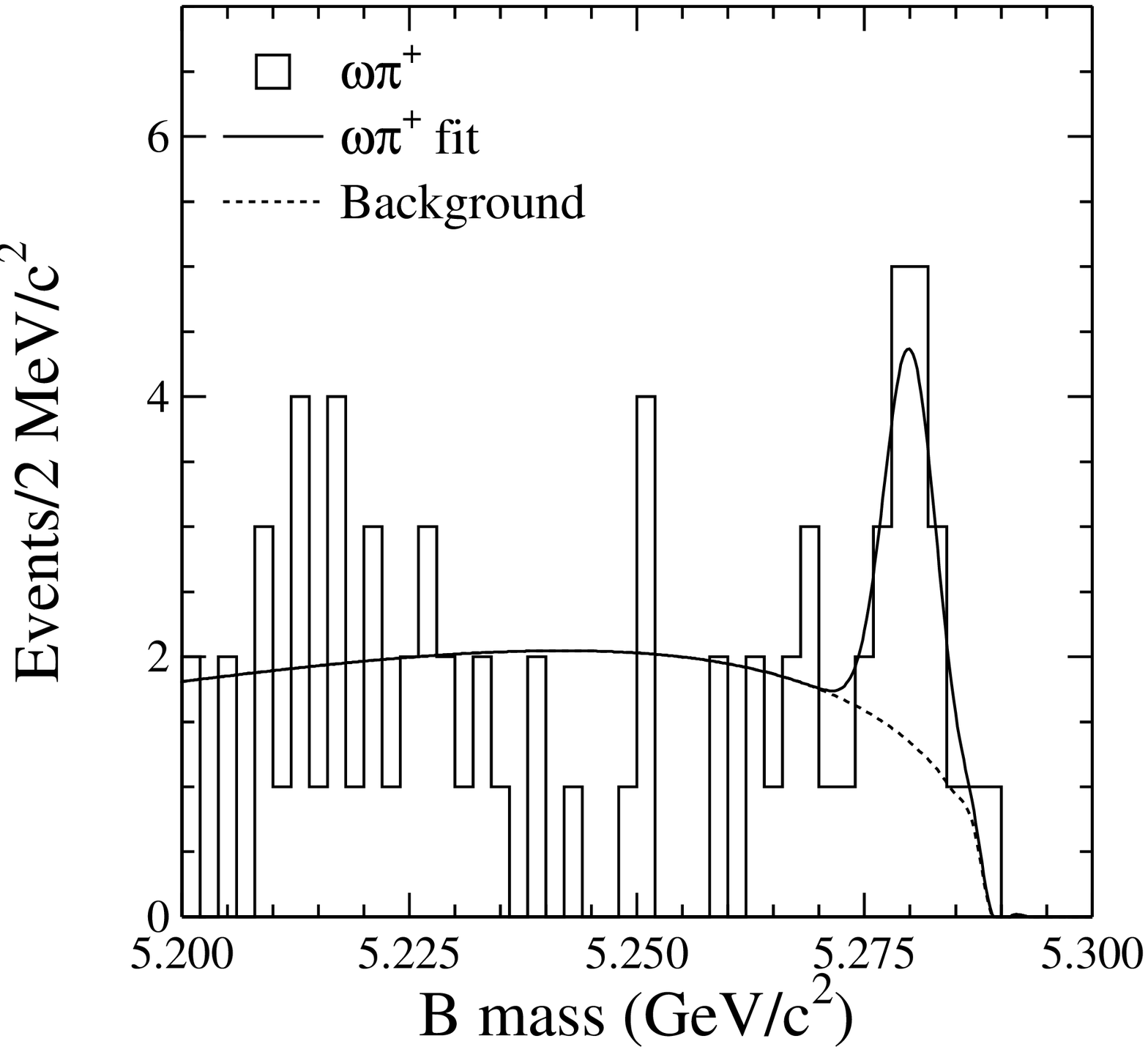}
\epsfxsize=3.0in
\epsfysize=3.0in
\epsffile{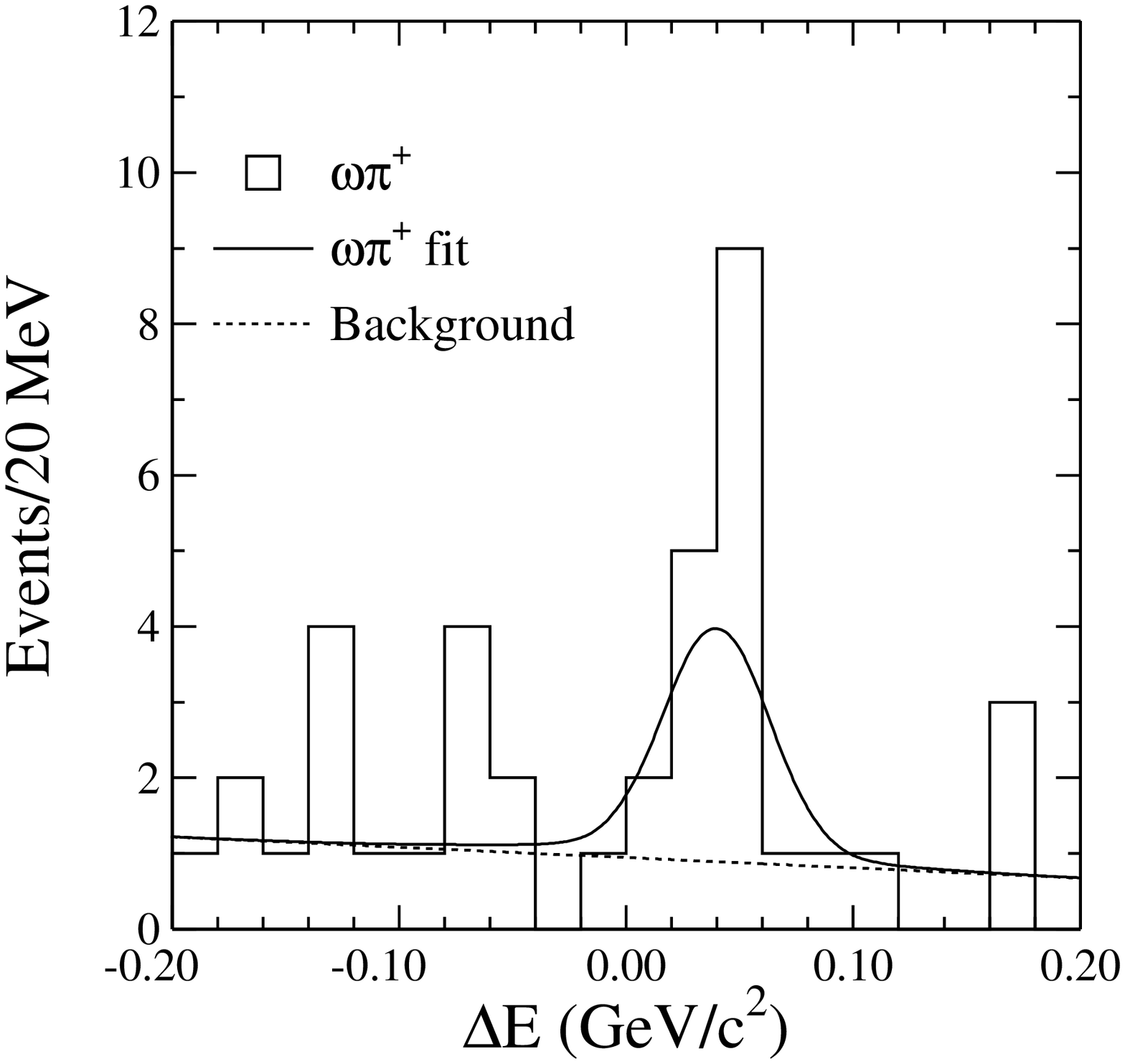}
 \caption{\label{f:proj1}
 Projection onto the reconstructed $B$ mass (left) and $\Delta E$ (right)
for \Bomegapi. The solid line shows the result of the likelihood
fit, scaled to take into account the cuts applied to variables not
shown. The dashed line shows the background component.}
\end{figure}

\begin{figure}[htbp]
\centering
\leavevmode
\epsfxsize=3.0in
\epsfysize=3.0in
\epsffile{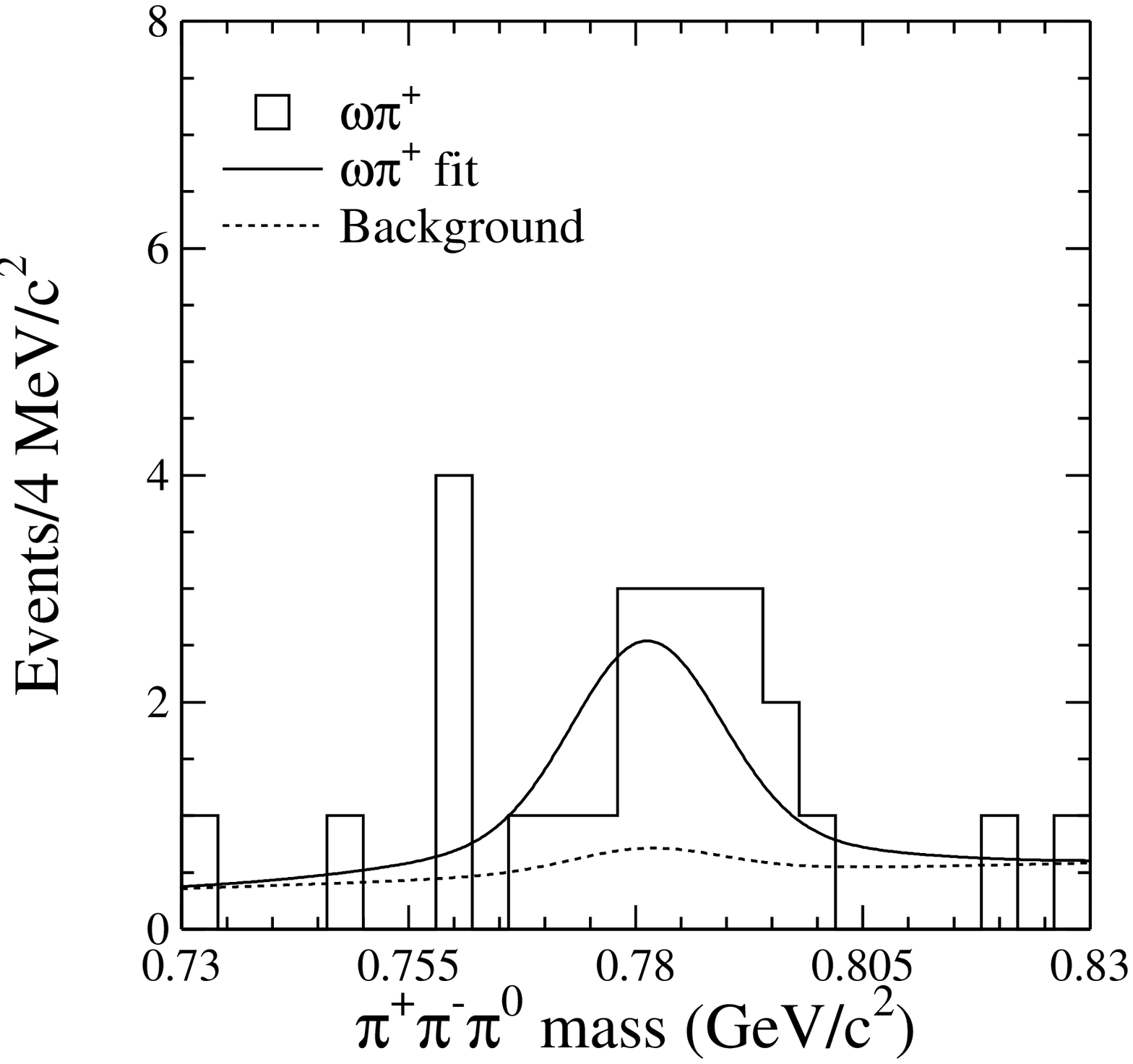}
\epsfxsize=3.0in
\epsfysize=3.0in
\epsffile{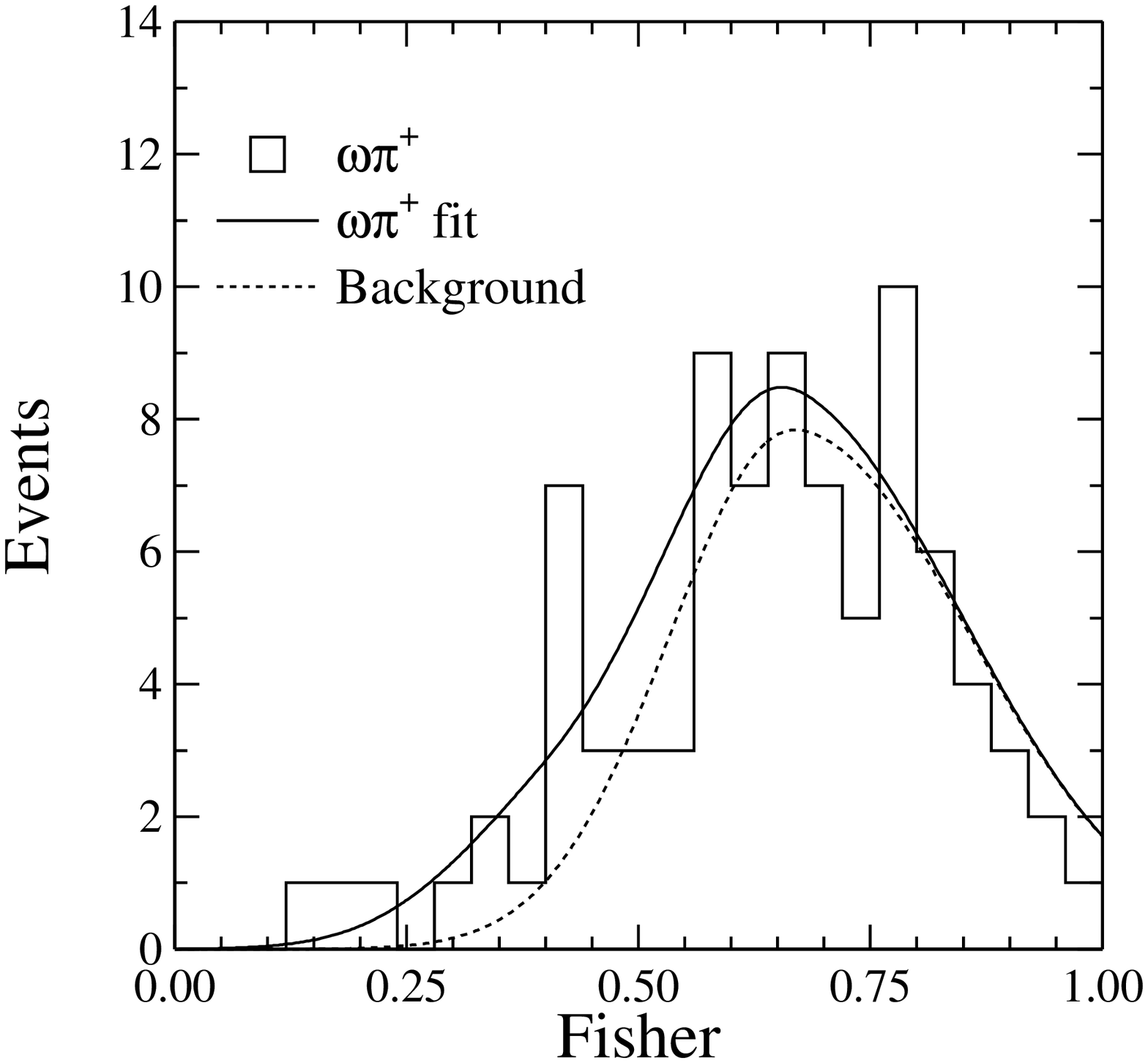}
 \caption{\label{f:proj2}
 Projection onto the reconstructed $\omega$ mass (left) and $\cal F$ (right)
for \Bomegapi. The solid line shows the result of the likelihood
fit, scaled to take into account the cuts applied to variables not
shown. The dashed line shows the background component.}
\end{figure}

 As can be seen in Table~\ref{results1}, we observe a clear signal for
the decay \Bomegapi. Figure~\ref{f:cont} shows the contours in
$1\sigma$ intervals of the likelihood function. Figures~\ref{f:proj1}
and \ref{f:proj2} show projections of the data on fit variables after
tight selection cuts were applied on the variables other than the one
plotted. The solid curves are not a fit to the data showed in each
plot, but rather an overlay of the fit function scaled to take into
account the additional cuts applied.

\subsection{The decays \Brhozpiz\ and \Bkstzpiz.}

 The results for the decay modes \Brhozpiz\ and \Bkstzpiz\ are based on a
 sample of $8.3 \times 10^6$ \bbbar\ pairs. There is significant cross-feed
 between the two decay modes, due in large part to the presence of a high
 momentum $\pi^0$ in the final state, which results in poorer \DE\ resolution.
 As a first step both decay modes are fit simultaneously. In this case, event
 selection prior to the maximum likelihood fit requires the invariant mass of
 the resonance to be within $0.3 < M_{h^+h^-} < 1.0$ GeV/$c^2$, as well as
 $-0.3 < \DE < 0.2$ GeV. Both quantities are computed assuming that the two
 charged tracks are pions. The continuum background hypothesis in the ML fit
 is separated into four components, depending on the identity of the charged
 tracks ($\pi^+$ or $K^+$). For input to the ML fit, \DE, the resonance mass,
 and $\cal H$ are calculated assuming the two charged tracks are pions for the
\rhozpiz\ signal and continuum background hypotheses, and
assuming one charged track to be a kaon and the other a pion for the \kstzpiz\
hypothesis. We also use the $dE/dx$ information for the two charged
tracks. The results of this combined fit established that there is no
significant signal for the decay mode \Brhozpiz. For this mode, the combined
fit results are reported in Tables~\ref{results1} and \ref{results2}.

 A second ML fit was then performed for \Bkstzpiz\ only. The preliminary
 selection required the $dE/dx$ information for both the charged pion and
 charged kaon candidates in the final state to be within $2\sigma$ of the
 expected value. We also required $|\DE| < 0.2$ GeV, and that the resonance
 mass be within 150~MeV/$c^2$ of the known $K^{*0}$ mass. The ML fit looked
 for only two hypotheses, signal \kstzpiz\ and continuum background. In this
 fit, no $dE/dx$ information was used as input. No significant yield was
 observed, as indicated in Table~\ref{results1}.


\section{DISCUSSION}

The observation of the decay \Bomegapi\ yields a branching fraction in
complete agreement with the previously reported measurement for \Brhozpi. For
the corresponding neutral decay modes \Bomegapiz\ and \Brhozpiz, fairly
stringent upper limits are set, as could be expected because of their color
suppression relative to the charged decay modes. For the latter decay mode,
the upper limit from this work is another piece of information for studies of
decays of the type $B \ra \rho \pi$, to go along with the previous
observations of
\Brhozpi\ and \Brhompi~\cite{aps99}.

We also see evidence for the decay
mode \Bomegakz. In this case the charged decay mode \Bomegak\ is expected to
have a similar branching fraction. For this latter mode, the
additional data and re-analysis of old data did not support the previously
reported observation~\cite{omega}. However, the central value for \Bomegak\
is only about $1.5\sigma$ away from the central value for \Bomegakz. There is
no significant evidence for decays of the type $B \ra \rho K$, although the
results are consistent with the ones for $B \ra \omega K$ decays.

The analysis presented here finds no significant yield for the decay mode
\Bkstzpiz, in agreement with theoretical expectations which point to a
branching fraction 
smaller than the previously observed \Bkstppi~\cite{aps99}. There is no
evidence for decay modes of the type $B \ra K^* K$, as well as for $\phi K$
and $\phi \pi$ final states.


\section{CONCLUSIONS}

We have observed the decay mode \Bomegapi, and measure a preliminary branching
fraction of ${\cal B}(\Bomegapi) = (11.3^{+3.3}_{-2.9} \pm 1.5) \times
10^{-6}$, in agreement with the previously reported observation of
\Brhozpi. We also see evidence for the decay \Bomegakz. New upper limits
are set for other final states including an $\omega$ meson, and for the
decays \Brhozpiz\ and \Bkstzpiz. In combination with previously reported
measurements and upper limits, these results are generally in agreement with 
predictions based on factorization models in charmless hadronic $B$ decays.

We gratefully acknowledge the effort of the CESR staff in providing us with
excellent luminosity and running conditions.  J.R. Patterson and
I.P.J. Shipsey thank the NYI program of the NSF, M. Selen thanks the PFF
program of the NSF, M. Selen and H. Yamamoto thank the OJI program of DOE,
J.R. Patterson, K. Honscheid, M. Selen and V. Sharma thank the A.P. Sloan
Foundation, M. Selen and V. Sharma thank the Research Corporation, F. Blanc
thanks the Swiss National Science Foundation, and H. Schwarthoff and E. von
Toerne thank the Alexander von Humboldt Stiftung for support.  This work was
supported by the National Science Foundation, the U.S. Department of Energy,
and the Natural Sciences and Engineering Research Council of Canada.



\begin{references}
\bibitem{conj} Charge conjugate modes are implied throughout this paper.
\end{references}
\end{document}